\documentclass[twocolumn,aps,prb,floatsfix]{revtex4}
\usepackage{graphicx}
\usepackage{amssymb,amsmath,mathtools}
\usepackage{array}
\usepackage{dcolumn}

\makeatletter
\renewcommand{\section}{\@startsection{section}{1}{-10pt}{-1.2\baselineskip}{.2\baselineskip}{\sffamily\bfseries}}
\makeatother  

\begin{document} 
\title{Understanding interface effects in perovskite thin films}

\author{Marie-Bernadette Lepetit}
\author{Bernard Mercey}
\author{Charles Simon}

\affiliation{CRISMAT, ENSICAEN-CNRS UMR6508, 6~bd. Mar\'echal Juin, 14050 Caen, 
FRANCE}

\date{\today}

\begin{abstract} 
  The control of matter properties (transport, magnetic, dielectric,\dots)
  using synthesis as thin films is strongly hindered by the lack of reliable
  theories, able to guide the design of new systems, through the understanding
  of the interface effects and of the way the substrate constraints are
  imposed to the material. The present paper analyses the energetic
  contributions at the interfaces, and proposes a model describing the
  microscopic mechanisms governing the interactions at an epitaxial interface
  between a manganite and another transition metal oxide in perovskite
  structure (as for instance $\rm SrTiO_3$). The model is  checked against
  experimental results and literature analysis.
\end{abstract}

\maketitle                                                                      

The technological importance of spin valves or spin injectors as potential
applications of manganese oxides induced a large number of works on manganite
thin films~\cite{EOM00,SCHLOM09}. For this reason the lost of magnetization of
$\rm La_{2/3}Sr_{1/3}MnO_3$ (LSMO) or $\rm La_{2/3}Ca_{1/3}MnO_3$ (LCMO) near
an $\rm SrTiO_3$ (STO) interface has been the subject of many interpretations.
Let us cite (i) homogeneous substrate strain~\cite{Ziese} (ii) electronic
and/or chemical phase separation~\cite{Fontcuberta} related to structural
inhomogeneities at the interface~\cite{BIBES}, (iii) manganese $e_g$ orbital
reconstruction 
inducing C-type antiferromagnetism~\cite{TEBANO_LD,TEBANO_ARPES}. None of
these interpretations however provide a good understanding of the observed
phenomena.  For instance, it was shown that an homogeneous substrate strain of
the in-plane parameters does not relax for film thickness smaller than
1000\AA{}~\cite{Fontcuberta}, while a drastic change in the transport
properties is observed for films thinner than a few unit cells ($\sim 3-4$ on
STO substrate~\cite{RAME,TEBANO_LD}, $\sim 30$ on $\rm LaAlO_3$
substrate~\cite{TEBANO_LD},\dots). In the second hypothesis (ii), there is no
clear proposition of the nature of the inhomogeneities, their origin, the way
they may act in order to induce the observed properties. Finally,
ferromagnetic hysteresis loops were found in very thin films up to only three
unit cells~\cite{RAME} (u.c.), in contradiction with the proposed C-type AFM ordering
resulting from orbital ordering (iii).  In any case, whatever the reasons put
forward, the existence of a so-called {\em ``dead layer''} at the interface
between the manganite film and most perovskite substrates seems to be
established~\cite{YAMAD,RAME}. This {\em ``dead layer''} is of a few unit
cells width and exhibits a large decrease of the conductivity~; however its
origin is not at all understood.

We believe that a careful analysis 
allow us to infer a model for the interface effects between a manganite and
an oxide substrate with a perovskite structure. The main concepts of our model
can be summarized as an energy balance at the interface.
\begin{itemize} 
\item It is well known that the strongest effect of the substrate is to
  constrain the film in-plane cell parameters to fit the substrate ones.
  \centerline{    $a_\text{film} = a_\text{substrate} \qquad b_\text{film} = b_\text{substrate}$}
  This constraint is quite strong since it is associated with bond elongation,
  i.e.  the most energetic vibrational modes~\cite{ABRASHEV}.  It thus relaxes
  slowly 
  (not before 250 u.c., 1000\AA{}, on a STO substrate~\cite{Fontcuberta}).  In
  the literature, it is associated with an u.c. volume constraint
  $V_\text{film}\simeq V_\text{bulk}$.
  While there is indeed, in the film free energy, an elastic term favoring
  $V_\text{film}=V_\text{bulk}$~:
  $\frac{V}{2\kappa}\,\left(\frac{\Delta V}{V}\right)^2$, this term cannot not
  be treated as a constraint imposed by the substrate.
  It should rather be evaluated against the other terms of the constrained
  film free energy.

\item The substrate imposes to the film its in-plane symmetry operations. This
  constraint is usually weak since, acting on bond angles and dihedral angles,
  it is related to low energy vibrational modes. One can thus expects that,
  after constraining the few first cells at the interface, these constraints
  start to relax according to the energy of the associated vibrational mode.

\item The electronic structure of the film and substrate interact at the
  interface. In particular the possible delocalization effects at the
  interface should be taken into account.  
\end{itemize}

\begin{figure}[h]
\begin{flushleft} (a) \end{flushleft} \hfill \\[-3eM] \resizebox{5.0cm}{!}{\includegraphics{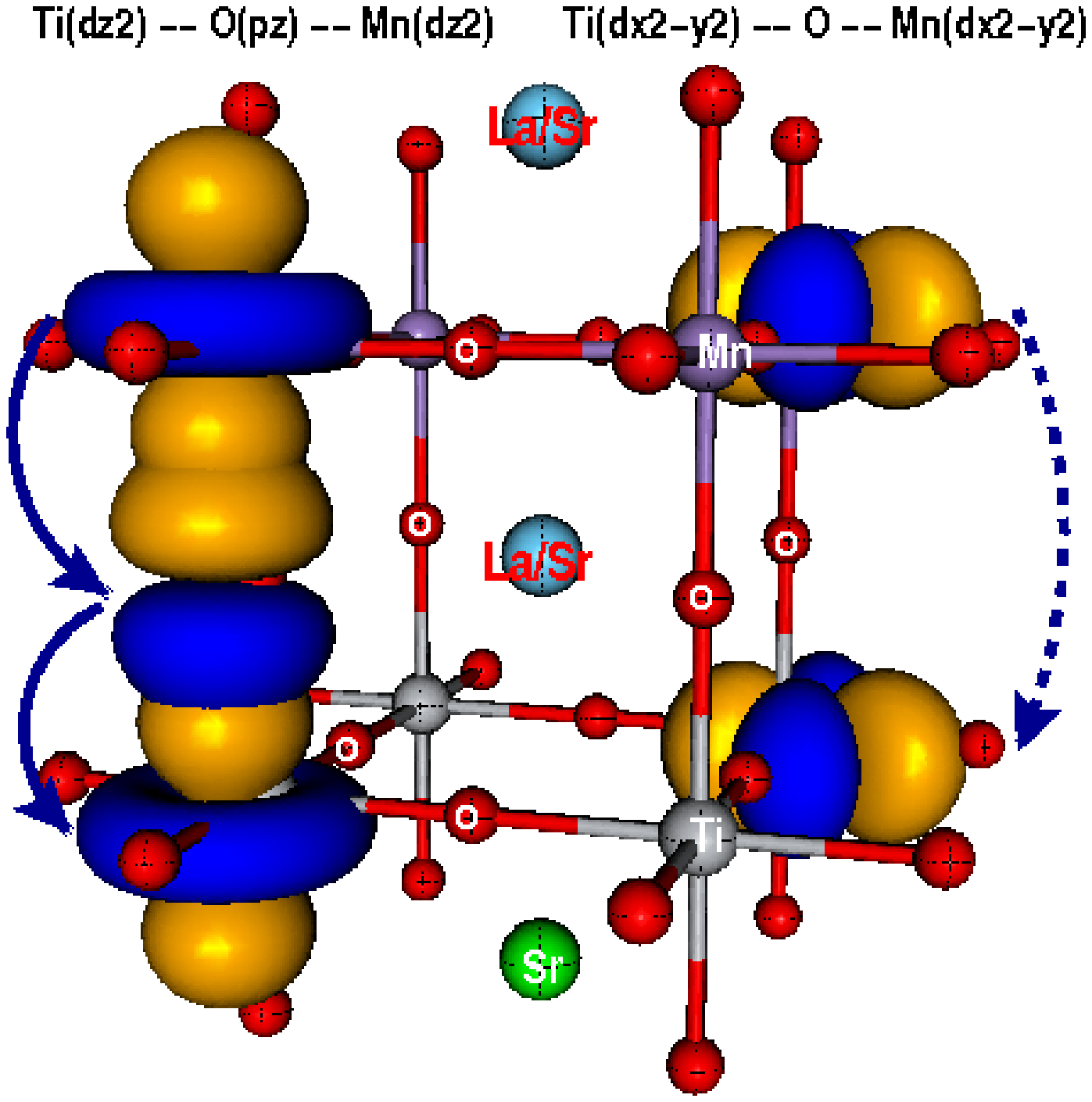}} \\[-1eM]
\begin{flushleft} (b) \end{flushleft} \hfill \\[-2eM] \resizebox{7.5cm}{!}{\includegraphics{t2.eps}}\\
\caption{(a) $d_{z^2}^\text{Mn}$---$2p_z^\text{O}$---$d_{z^2}^\text{Ti}$ and  
$d_{x^2-y^2}^\text{Mn}$---(O)---$d_{x^2-y^2}^\text{Ti}$ delocalization mechanism. (b)
Through bridge delocalization mechanisms. The resulting effective
  Mn---Ti transfer integral is at the second order of perturbation \\
  $\scriptstyle t_{d_{z^2}^\text{Mn},d_{z^2}^\text{Ti}} = - \;
  \langle d_{z^2}^\text{Mn}|\hat H|p_z^\text{O} \rangle 
  \langle{\rm O} p_z^\text{O}|\hat H| d_{z^2}^\text{Ti}\rangle
  \;/\;(\varepsilon_d - \varepsilon_p)$.}
\label{fig:dpd}
\end{figure}

As a matter of example let us see how these constraints apply to a LSMO thin
film on an STO (001) substrate. The in-plane parameters of the LSMO are
imposed by the STO substrate and the film in under tensile strain at the
interface. As already mentioned, this constraint holds over a large
number of monolayers (ML).
%
The STO in-plane symmetry operations impose to the first layers of LSMO 
to present a 4-fold symmetry axis, perpendicular to the film, and untilted
octahedra. It results, that in these layers, the crystal-field split $3d$
orbitals of the Mn atom share their $(\vec x,\vec y,\vec z)$ orthogonal axes
with the $(\vec a,\vec b,\vec c)$ substrate lattice vectors ($\vec c$ being
the out-of-plane direction)~; that is the Mn orbitals are $d_{ab}$, $d_{ac}$,
$d_{bc}$ for the low energy ones and $d_{3c^2-r^2}$, $d_{a^2-b^2}$ for the
high energy ones. Any further distortion (such as Jahn-Teller) of the $\rm
MnO_6$ octahedra should thus respect those constraints and are therefore
restricted to atomic movements along the $\vec c$ direction.
Let us now analyze the energy minimization of the LSMO layers at the interface
under the above constraints. On one hand the film is under tensile strain and
the minimization of the elastic energy favors a contraction of the LSMO
monolayer u.c. along the $\vec c$ axis. On the other hand the film Fermi level
orbitals are the Mn partially filled $d_{z^2}$ and $d_{x^2-y^2}$ ones while
the lowest empty orbitals of the STO are the Ti empty $3d$ orbitals.  It is
thus quite natural that the $e_g$ manganese orbitals delocalize to some extend
into the titanium $3d$ empty ones. This interaction is particularly favored in
the case of the $3d_{z^2}$ orbitals of the Mn and Ti atoms (see
figure~\ref{fig:dpd}).
Indeed, the through-bridge bonding mechanism 
acting via the $2p_z^\text{O}$ orbitals of the oxygens is energetically favored, due to
the large $d_{z^2}^\text{Mn}$---$2p_z^\text{O}$---$d_{z^2}^\text{Ti}$ overlaps
(fig.~\ref{fig:dpd}). This delocalization energy tends to favor a Jahn-Teller
distortion increasing the occupation of the $d_{z^2}^\text{Mn}$ orbitals, that is an
elongation of the LSMO monolayer u.c.  along the $\vec c$ axis.
The physics at the interface thus results from the competition between 
\begin{itemize}
\item the elastic energy that favors $c<a$, \vspace*{-1.5ex}
  $$\frac{V}{2\kappa}\,\left(\frac{\Delta V}{V}\right)^2$$ 
\item and the delocalization energy that favors  $c>a$  \vspace*{-1.5ex}
$$-\,\frac{(t_{d_{z^2}^\text{Mn},d_{z^2}\text{Ti}})^2}{\varepsilon_{d_{z^2}^\text{Ti}}
    - \varepsilon_{d_{z^2}^\text{Mn}}} \vspace*{-1.5ex}$$ 
$$\text{where }\quad t_{d_{z^2}^\text{Mn},d_{z^2}^\text{Ti}} \simeq - \frac{\langle d_{z^2}^\text{Mn}| p_z^\text{O}\rangle
    \langle p_z^\text{O}| d_{z^2}^\text{Ti}\rangle}{\varepsilon_d
    - \varepsilon_p}$$ 
\end{itemize}
On one hand, the relatively short metal--oxygen distances ($\simeq
1.95\rm\AA{}$), the strong directionality of the $d_{z^2}$ and $p_z$ orbitals
inducing a very large $\langle d_{z^2}^\text{Mn}|p_z^\text{O}\rangle \langle
p_z^\text{O}| d_{z^2}^\text{Ti}\rangle$ overlap term, and the relatively weak
orbital energy difference between the oxygen $2p$ and the metal $3d$ orbitals
results in a very large effective transfer integral between the $d_{z^2}$
orbitals of the Mn and Ti atoms (of the order of $\simeq 1\rm
eV$~\cite{GUIHERY}) and thus of the delocalization energy. Indeed, one do not
expect the $\varepsilon_{d_{z^2}^\text{Ti}} - \varepsilon_{d_{z^2}^\text{Mn}}$
energy difference
to be larger than a few electron-Volts ($\sim$1,2).
On the other hand, the elastic energy can be evaluated for different $c$
values. For a cubic structure ($c=c_{STO}=3.905\rm\AA{}$), the LSMO volume
increase is of 2\%, and the associated elastic energy can be estimated to
$\simeq 0.015\,\rm eV$ (the LSMO compressibility being taken from
ref.~\onlinecite{GLASSER}), that is much weaker (an order of magnitude) than
the expected delocalization energy. Even if these energetic evaluations are
only qualitative, one can expect with quite confidence, that for LSMO films
(LCMO and similar films) on a STO substrate the first few ML at
the interface are elongated in the $\vec c$ direction, despite their already
in-plane tensile strain.

After a few unit cells, there is no more delocalization energy to gain from a
larger occupation of the $d_{z^2}$ orbitals. At the same time, the in-plane
symmetry operations imposed by the substrate should start to relax, and in
particular the ones associated with the LSMO vibrational modes with the lowest
frequencies~: the octahedra tilt. The consequence of this constraint relaxation
should be to adjust the $e_g$ orbitals occupations toward a  bulk-like value 
(compared to the first interface ML  it means an increase of
the $d_{x^2-y^2}$ occupation and a decrease of the $d_{z^2}$ one). The
 $c$ ML parameter should thus start to decrease toward the value
expected from the elastic energy minimization (monolayer u.c. volume
conservation), that is a contraction of the monolayers $c$ parameter and a
tendency to a larger occupation of the $d_{x^2-y^2}$ orbitals compared to the
$d_{z^2}$ ones.


What are the consequences of the above structural considerations, in term of
magnetic and transport properties? As far as the first few layers at the
interface are concerned, the tendency to occupy the $d_{z^2}$ orbitals at the
expense of the $d_{x^2-y^2}$ ones should result in a decrease of the
(in-plane) double-exchange and thus in a strong reduction of both the Curie
temperature and the conductivity~\cite{DeGennes}. Let us note that even the
$d_{z^2}$ orbitals are subject to the (in-plane) double-exchange, however the
$d_{z^2}^\text{Mn}$---$p_{x/y}^\text{O}$---$d_{z^2}^\text{Mn}$ delocalization
process is much less effective than the
$d_{x^2-y^2}^\text{Mn}$---$p_{x/y}^\text{O}$---$d_{x^2-y^2}^\text{Mn}$ one,
with the consequence that the effective (in-plane) exchange integral is much
weaker and thus the Curie temperature. After a few layers, the relaxation of
the $e_g$ orbitals occupation should result in an increase of the double
exchange and thus of both the Curie temperature and the film conductivity. The
temporary limit being set by the in-plane $a,b$ parameters constraints.

Finally, for very thick films this later constraint is relaxed and
one should retrieve the bulk properties.

Let us now check the predictions issued from our model against the
experimental data. LSMO thin films were grown by Laser-MBE using a
setting detailed in ref.~\onlinecite{SALVA}. 
The temperature during the deposition is 620$^\circ$C and the pressure is $4
\times 10^{-4}$mbar. The gas is a mixture of $\rm O_2$ and $\rm O_3$.
The intensity oscillations of the specular beam (RHEED) are used to measure
the number of deposited layers. After the deposition, the pressure is
increased to $5 \times 10^{-3}$mbar associated with a higher $\rm O_3$
concentration, to insure a good oxidation. The films are then
cooled to room temperature in this atmosphere. Such deposition conditions
allow us to grow films with a $T_c$
higher than 325\,K for thicknesses larger than 100\AA{}.
%
After the growth, the films were studied by X-ray diffraction
($\theta-2\theta$ mode) Seifert XP3000 system. From these data, the $c$
lattice parameter and the films thicknesses were determined.
These measured thicknesses agree well with the number of RHEED oscillations.
%
Magnetization measurements were carried out in a Quantum Design SQUID
system, 
under a 500~Oersteds magnetic field parallel to 
the substrate. 
Transport measurements, four probes method, were carried out in a PPMS Quantum
Design system.

LSMO ultrathin films with thicknesses ranging from 31\,\AA{} (8 ML) to
92\,\AA{} (24 ML) were grown 
onto 001-oriented $\rm SrTiO_3$ substrates, previously etched to ensure a $\rm
TiO_2$ terminating layer~\cite{KAWA}.  The out-of-plane average ML 
lattices parameters, $\langle c \rangle$~, of the different films were
determined by X-ray diffraction and are reported in table~\ref{tab:c}.
\begin{table}[h]
\begin{tabular}{c@{\hspace{3ex}}c@{\hspace{3ex}}c@{\hspace{3ex}}D{.}{.}{2}@{\hspace{3ex}}}
\hline
Thickness (\AA{}) & Monolayers & $\langle c\rangle$ (\AA{}) &  \multicolumn{1}{c}{$\langle
V\rangle$ ($\rm \AA{}^3$)}\\
\hline
\multicolumn{2}{c}{Expected values} & 3.823 & 58.29~\cite{VENKA} \\
$>$ 200 & $>$ 52 & 3.847 & 58.66\\
92 & 24 & 3.854 & 58.77 \\
70 & 18 & 3.856 & 58.80 \\
54 & 14 & 3.863 & 58.91 \\
31 & 8 & 3.874  & 59.07 \\
\hline
\end{tabular}
\caption{Film thicknesses (\AA{}), numbers of LSMO monolayers, average
  monolayer $c$ parameters and average ML unit cell volume 
  for  LSMO films on STO (001) substrate.}
\label{tab:c}
\end{table}
One sees immediately that the variation of $\langle c\rangle$ as a function of
the number of ML 
does not correspond to the value expected from the conservation of the
manganite volume. Indeed, the LSMO tensile strain at the interface let us
expect a reduction of the ML average $\langle c \rangle$ parameter compared to
the bulk value. Table~\ref{tab:c} exhibits large $\langle c\rangle$ values
associated with ML volumes larger than $V_\text{bulk}$.  Such a behavior was
also observed by Herger \&~al~\cite{HERGER} on even thinner films with only
1,2,3,4,6 and 9 ML of LSMO.  Our films show an increase of $\langle c\rangle$
when reducing the thickness (in agreement with films of
ref.~\onlinecite{HERGER}). Further examining the data of
reference~\onlinecite{HERGER}, one sees that while the first few layers at the
interface exhibit $c$ values larger than the STO one (cumulative displacement
compared to $c_\text{\tiny STO}$~: $\Delta z$ positive and increasing), after
about 3 monolayers $c$  retrieves a value smaller than
$a=a_\text{\tiny STO}$ (decreasing $\Delta z$). At this point, let us notice
that all these structural results are in full agreement with our theoretical
predictions.

\begin{figure}[h]
\resizebox{8cm}{!}{\includegraphics{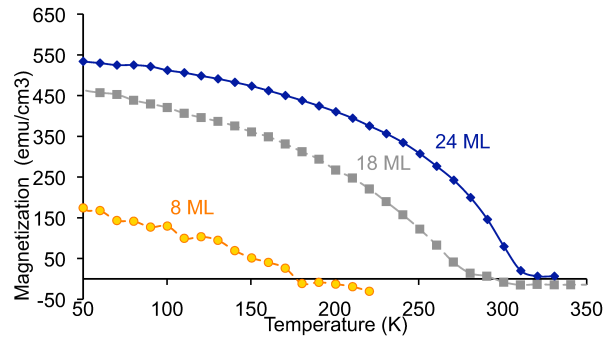}} \vspace*{-1.5eM}
\caption{Magnetization under a 500~Oersteds field of LSMO films on a STO (001)
  substrate.}
\label{fig:films}
\end{figure}

Let us now examine the magnetic and transport properties.  The
magnetization of the different films is reported 
on figure~\ref{fig:films}.
For films thicker than 14 ML, the magnetization remains close to the
values observed for the thickest films and the $T_c$ remains
high. However, when the number of ML is decreased to 8, the
magnetization at 10\,K as well as $T_c$ are strongly
decreased.  These results agree well with the values reported in other
works~\cite{RAME,TEBANO_LD} for ultrathin LSMO films.
The resistivity measurements agree with the suppression of the double exchange
suggested for 8 ML films on the magnetization curves
(fig.~\ref{fig:films}). Indeed, the three thickest samples (14, 18 and 24 ML)
exhibit a metal-insulator transition at $T_c$ while the thinner one (8 ML)
presents a semiconductor-like behavior with a large magneto-resistive effect.
These peculiar magnetic and transport properties of the first few ($\sim$ 6-8)
LSMO layers at the interface with STO are usually referred at as {\em the dead
  layer}~\cite{RAME,TEBANO_LD}.

At this point let us notice that these experimental results do agree nicely
with our theoretical predictions, and allow us to determine the number of
layers over which each constraint extends. For instance the dominating
$d_{z^2}$ occupation can be associated with only the first 2-3 ML (see $T_c$
or $\Delta z$ variation in the 9 ML film of ref.~\onlinecite{HERGER}). In the
next layers the $e_g^\text{Mn}$ orbitals occupations start to relax toward a
more balanced one as supported by the increase of both $T_c$ and conductivity
as a function of the number of layers in 3 to 8 ML films. The fact that the
$T_c$ value reaches its saturation limit, that the metal-insulator transition
is restored tells us that after about 8 ML, the double-exchange is fully
restored and thus the manganese $e_g$ orbitals occupations.

The last point of our predictions one would like to see verified, is the
actual dominating $d_{z^2}^\text{Mn}$ occupation in the first layers at the
STO interface. LD-XAS experiments were conducted by different authors that
reached opposite conclusions. Indeed, while Tebano {\it
  et~al}~\cite{TEBANO_LD} report the signature of a preferential $d_{z^2}$
orbital occupation at the interface, Huijben {\it et~al} argue that the
experimental evidences are not significant. Very recently ARPES
measurements~\cite{TEBANO_ARPES} were done on 4,6 and 10 ML films and clearly
exhibit an increase of the $\Gamma$ point signal (attributed to the
$d_{z^2}^\text{Mn}$ orbitals contribution) when the thickness of the film
decreases. It thus seem that the experimental data confirm our findings from
constraints and energetic considerations.

Another way to check the validity of our interface model is to test it against
the properties of ultrathin films deposited on a buffered substrate. Indeed, the
deposition of a buffer layer at the interface between the STO and the LSMO
should strongly modify the electronic structure of the first LSMO layers.  We
thus grew either a $\rm (LaMnO_3)(SrMnO_3)$ buffer (BUF1) or a $\rm
[(LaMnO_3)(SrMnO_3)]_2$ (BUF2) on STO, prior to the LSMO film. Let us point
out that the $\rm (LaMnO_3)(SrMnO_3)$ superlattices present an
antiferromagnetic behavior~\cite{VERBEECK}, when deposited in these
conditions.  The buffer layer will thus break both the symmetry constraint
preventing the LSMO octahedra to be tilted in the first LSMO layers at the
interface and the possibility for the LSMO manganese to delocalize its
$d_{z^2}$ population in the empty $d_{z^2}^\text{Ti}$ orbitals. Of course some
delocalization of the LSMO $d_{z^2}^\text{Mn}$ orbitals can still occur since the
buffers $d_{z^2}^\text{Mn}$ orbitals are most probably not totally
filled. This effect is however expected to be much weaker than at the LSMO/STO
interface both because of the buffer  $d_{z^2}^\text{Mn}$ orbital occupation
and because of the starting of the $\rm MnO_6$ octahedra tilt in the LSMO.  We
thus expect that the buffered LSMO films will be more ``bulk-like'' than the
unbuffered ones, the thicker the buffer, the more ``bulk-like'' the film (the
wider the distance between the film  and the STO surface, the
lesser the symmetry constraints will be imposed to the LSMO) . It means that
for equal film thicknesses, the thicker the buffer,  the highest the
$T_c$ and the film conductivity should be.

\begin{figure}[h]
\resizebox{8cm}{!}{\includegraphics{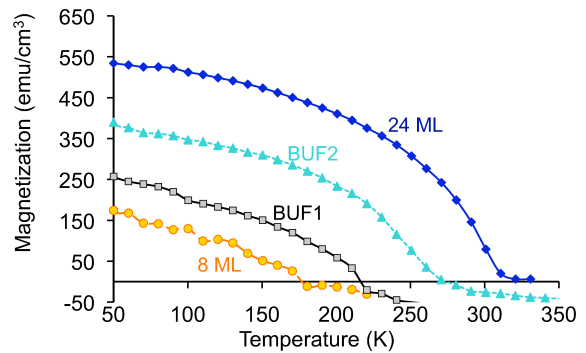}}
\caption{ Magnetization under a 500~Oersteds field, of 8 ML LSMO films
  deposited on bare or buffered (BUF1\,: $\rm (LMO)(SMO) \times 1$~; BUF2\,:
  $\rm (LMO)(SMO) \times 2)$ $\rm SrTiO_3$ substrates. The magnetization of an
  unbuffered 24 ML film is given for comparison.}
\label{fig:buff}
\end{figure}

\begin{table}[h]
\begin{tabular}{c@{\hspace{3ex}}D{.}{.}{2}@{\hspace{3ex}}}
\hline
Substrate & \multicolumn{1}{c}{$\langle c \rangle$ (\AA{})} \\
\hline
Bare & 3.874\\
Buffered $\rm (LMO)(SMO) \times 1$ & 3.871 \\
Buffered $\rm (LMO)(SMO) \times 2$ & 3.862 \\
\hline
\end{tabular}
\caption{Average   $\langle c \rangle$ ML parameters for 
8 ML LSMO films deposited either on bare or on buffered
substrates.}
\label{tab:param2}
\end{table}
Table~\ref{tab:param2} experimental data show that the average $\langle c
\rangle$ ML parameter decreases when the number of buffer layers goes from 0
to 2. This results in a decrease of the $c/a$ ratio and induces a
larger occupation of the $d_{x^2-y^2}$ orbitals with respect to the $d_{z^2}$
ones. Such an electronic structure favors a more robust ferromagnetic behavior
of the films as the thickness of the buffer layer increases.
Indeed, both the magnetization (see fig.~\ref{fig:buff}) and $T_c$ increase
with increasing buffer thickness, as expected from the $c/a$ ratio. The film
buffered by two (LMO)(SMO) layers exhibits a $T_c$ close to 275\,K, that is
about 100\,K higher than the $T_c$ observed for the film of same thickness
deposited on a bare STO substrate.
The resistivity of the buffered films behave as for a regular ferromagnetic
materials 
with a maximum value at $T_c$.

In conclusion, we presented in this paper a simple model able to explain and
predict the interface effects between perovskites oxides. We applied our model
to the difficult case of the nature and origin of the so-called dead layer at
the LSMO interface with an STO substrate. Our model successfully predicted and
explained all experimental features of this interface, up to now not
understood.

\acknowledgments The authors thank ANR program SEMOME and STREP MACOMUFI for
the support of this work.


\end{document}